\documentclass{acm_proc_article-sp}
\pdfoutput=1
\usepackage{gensymb}
\usepackage{url}
\usepackage{flushend}

\begin{document}

\toappear{This paper is in the public domain. It has been peer-reviewed and accepted at FPGAworld 2012, but ACM refuse to publish public domain papers from non-US government authors unless paid an unreasonable amount of money. ACM have decided to ignore the conference's scientific referee, disrespect their constitutional purpose of ``fostering the open interchange of information'', and remove this paper from the official conference proceedings on profit grounds.}

\title{A 26 ps RMS time-to-digital converter core for Spartan-6 FPGAs}
\numberofauthors{1}
\author{
\alignauthor
S\'ebastien Bourdeauducq\\
       \affaddr{Independent researcher}\\
       \email{sebastien.bourdeauducq@lekernel.net}
}

\maketitle
\begin{abstract}
We have designed, implemented and tested a time-to-digital converter core in a low-cost Spartan-6 FPGA. Our design exploits the finite propagation speed in carry chains to realize a delay line in which the propagation distance of the incoming signal's edges is measured using hundreds of taps. This technique enables the core to reach a precision far better than the minimum switching period of the FPGA flip-flops. To compensate for process, voltage and temperature (PVT) effects, our design uses a combination of two techniques: startup calibration and online calibration. The startup calibration uses a statistical method to estimate the delay between the taps of the delay line and helps eliminate the effect of process variations. The online calibration, which takes place without disruption of the core's operation, uses a ring oscillator whose frequency instability is measured and used to compensate for subsequent voltage and temperature effects on the delay line. Our tests show that our design reaches a precision of $26$ ps RMS over a temperature range of 37\degree C to 48\degree C.
\end{abstract}



\section{Introduction}
Several FPGA-based time-to-digital converter (TDC) designs have already been proposed\cite{epfl}\cite{fermilab}. However, there were many incentives for us to design a new core.

The PVT compensation mechanism of \cite{epfl} introduces dead times during which the core is insensitive to incoming signal transitions, which we found undesirable. The continuous calibration process of \cite{fermilab} requires that the statistical distribution of transitions within the reference TDC clock periods is uniform. This may not be the case in systems meant to be part of a particle accelerator, where many events are synchronous to a single clock. Therefore, we devised another technique (\textit{online calibration}) that does not introduce dead times and is independent of the statistics of the incoming signal.

We also wanted the design to function on Spartan-6 FPGAs so that it can be used on the SPEC\cite{ohwr} boards. Previous works are based on Virtex-5 or Cyclone-II FPGAs.

Finally, no source code is published for any of these designs, which renders it necessary to develop a new core for all practical purposes (and incidentally makes it more difficult to reproduce and verify the results). Our core is available under the LGPL free software license and its full VHDL code can be freely downloaded from \url{http://www.ohwr.org/projects/tdc-core}.

\section{Design}
\subsection{Overview}
\begin{figure*}[h!]
\centering
\includegraphics[width=\textwidth]{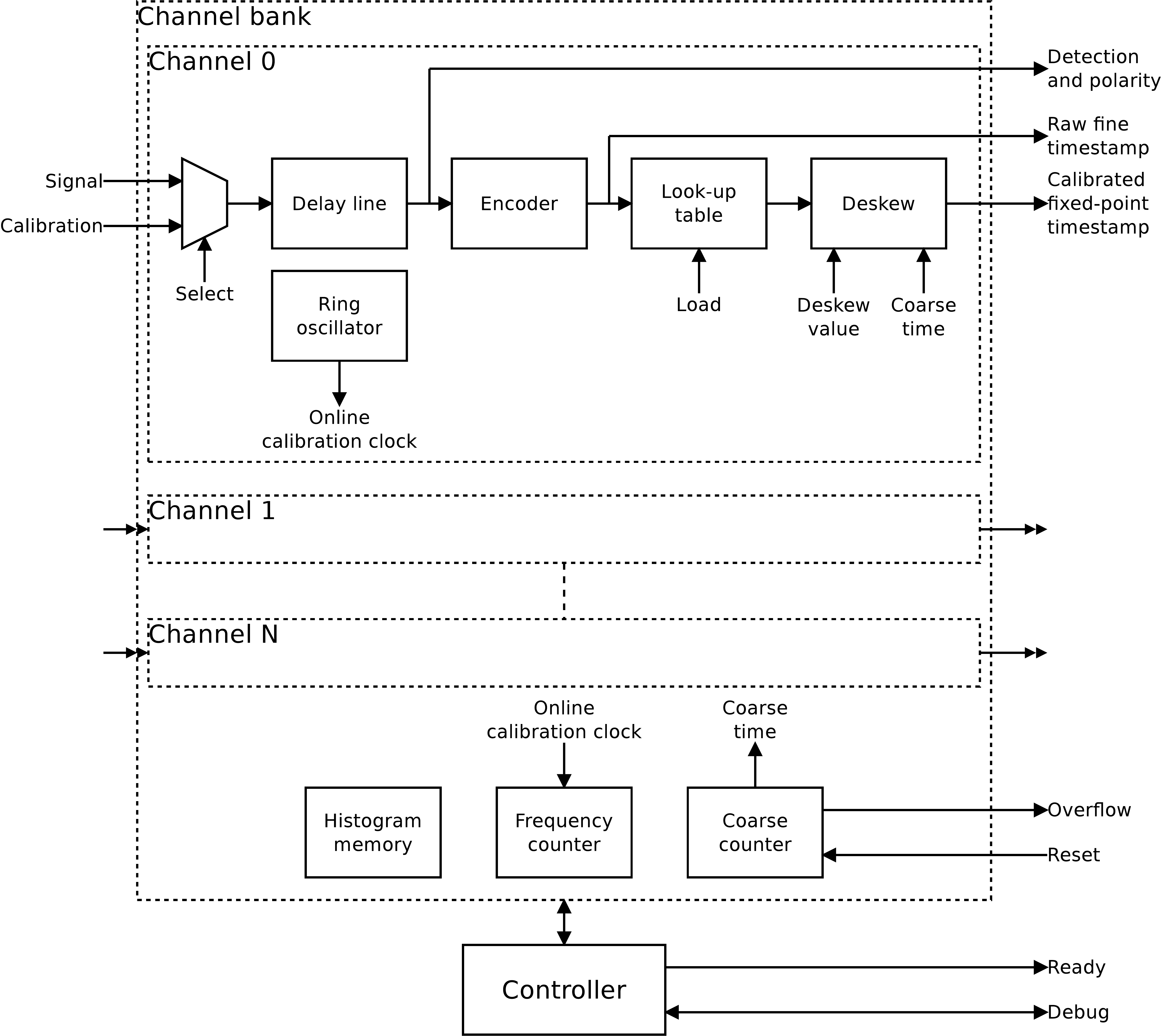}
\caption{Block diagram of the TDC core.}
\label{fig:block}
\end{figure*}

The block diagram of the core is given in Figure~\ref{fig:block}.

The generated timestamp is based on a cycle count and the arrival time within a clock cycle. The former needs only a simple counter whereas the latter is measured with a tapped delay line. The fine time measurement is obtained by injecting the signal into the tapped delay line which gives a measurement analogous to a thermometer after the taps are sampled by D flip-flops. The total delay of the delay line must be greater than the clock period. At each clock tick, an encoder counts the taps the signal has reached and gives a \textit{raw} measurement of the timestamp of the signal within the current clock cycle. This raw value is fed into a look-up table (LUT) which converts it into a calibrated value expressed in subdivisions of the clock cycle, called the \textit{fractional} value. Finally, in the \textit{deskew} stage, the fractional value is combined with the index of the current clock cycle given by the coarse counter, and the resulting fixed-point value is added a user-defined constant to enable the TDC core to directly generate timestamps relative to the source of the system clock.

The main difficulty with this system is that the delay line is subject to process, temperature and voltage (PVT) induced variations, and it needs to be calibrated against them.

To generate the LUT contents, the controller switches to the calibration signal. The key property of the calibration signal is that the probability density of its transition timestamps within a system clock cycle must be constant. The controller measures the raw timestamps and books a histogram. Because of the constant probability density, the heights of the histogram bars are approximately proportional to the delays between the taps of the delay line after enough measurements have been taken. Further, the last tap to have recorded a signal transition corresponds to a delay equal to the system clock period. This enables the controller to build the initial contents of the LUT. This process is called \textit{startup calibration}.

The drawback of the startup calibration is that the system cannot operate while the calibration is taking place. Therefore, a process of \textit{online calibration} has been devised. Each channel contains a ring oscillator that is placed close to the delay line. The controller periodically measures the frequency of this ring oscillator, compares it to the frequency that was measured at the time of startup calibration, linearly interpolates the fractional timestamps, and updates the LUT. This allows compensation of temperature and voltage effects while the system keeps running.

The system gives timestamps of both rising and falling edges of the incoming signal. The rising edges are discerned from the falling edges using the ``polarity'' output.

\subsection{Delay line structure}
\label{delaystruct}
The delay line uses a carry chain. It is made up of \verb!CARRY4! primitives whose \verb!CO! outputs are registered by the dedicated D flip flops of the same slices. The signal is injected at the \verb!CYINIT! pin at the bottom of the carry chain. The \verb!CARRY4! primitives have their \verb!S! inputs hardwired to 1, which means the carry chain becomes a delay line with the signal going unchanged through the \verb!MUXCY! elements (see \cite{s6hdl} for reference). Since each \verb!CARRY4! contains four \verb!MUXCY! elements, the delay line has four times as many taps as there are \verb!CARRY4! primitives.

Using the Xilinx timing model, a surprising observation is that some delay differences between consecutive taps are negative. This probably is at the origin of the ``bubbles'' mentioned in the EPFL paper \cite{epfl}. The schematics given by Xilinx of the \verb!CARRY4! primitive is misleading there, and has probably little to do with the actual transistor-level implementation. The Xilinx documentation \cite{s6hdl} gives a hint by describing the primitive as ``Fast Carry Logic \textit{with Look Ahead}''.

\begin{figure}[h]
\includegraphics[width=\columnwidth]{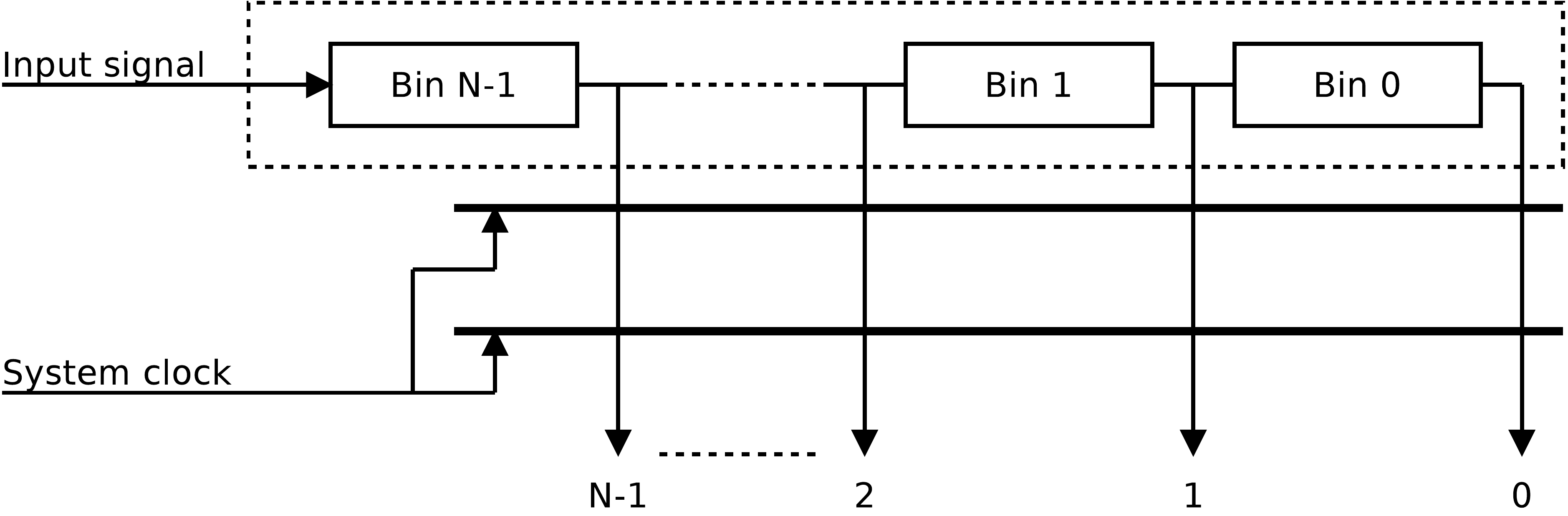}
\caption{Representation of the delay line.}
\label{fig:delaystruct}
\end{figure}

To avoid negative differences, we simply reorder the bits at the output of the delay line to sort the taps by increasing actual delays. We can then think of the delay line according to Figure~\ref{fig:delaystruct}. The bin widths are uneven, but the incoming signal reaches the taps in order. This last property simplifies the encoder design, since it only has to count the number of identical bits at the beginning of the delay line.

\subsection{Calibration details}
In the formulas below:
\begin{itemize}
\item $T_{sys}$ is the system clock period.
\item $H(n)$ is the number of hits in the histogram at output $n$. A hit at output $n$ means that the signal propagated down to output $n$, without reaching output $n-1$.
\item $W(n)$ is the width of bin $n$.
\item $C = \displaystyle\sum\limits_{n=0}^{N-1} H(n)$ is the total number of hits in the histogram.
\item $R(n)$ is the timestamp of an event whose signal propagated down to output $n$ (without reaching output $n-1$), measured backwards from the clock tick.
\item $f$ (respectively $f_{0}$) is the current (respectively reference) frequency of the online calibration ring oscillator.
\end{itemize}

\subsubsection{Offline calibration}
At startup, the core sends random pulses into the delay line (coming from a on-chip ring oscillator), builds the histogram, computes the delays (as explained in \cite{fermilab}), and initializes the LUT.

We take the first output of the delay line to be the origin of the time measurements, and we define:
\begin{equation}
W_{0}(N-1) = 0
\end{equation}

The width of other bins is proportional to their respective number of counts in the histogram. The widths sum up to a clock period. This leads to the following equation:
\begin{equation}
W_{0}(n) = \frac{H(n+1)}{C} \cdot T_{sys}
\end{equation}

The timestamp is the sum of the widths of the traversed bins:
\begin{equation}
R_{0}(n) = \displaystyle\sum\limits_{i=n}^{N-1}{W_{0}(i)} = \frac{T_{sys}}{C} \cdot \displaystyle\sum\limits_{i=n}^{N-1}{H(i)}
\end{equation}

In the TDC core, the unit is the clock period, and the output has $F$ base 2 digits after the radix points. The controller also chooses $C=2^{F+P}$, where $P$ is the number of \textit{extra histogram bits}. Expressed in units of $2^{-F}$ clock periods (which is the weight of the least significant bit of the fixed-point output), we have:
\begin{equation}
\frac{T_{sys}}{C}=2^{-P}
\end{equation}

\subsubsection{Online calibration}
Online calibration is performed with a simple linear interpolation of the delays relative to the ring oscillator frequencies:
\begin{equation}
R(n) = \frac{f_{0}}{f} \cdot R_{0}(n)
\end{equation}

Note that when $f < f_{0}$, some values can go above the maximum fractional part value of $1 - 2^{-F}$ and might not fit in the LUT anymore. However, those correspond to delays that now exceed one clock period, and therefore they should almost never get used. In case of overflow, the controller saturates the result by using the maximum value $1 - 2^{-F}$ in order to give the best approximation in case those LUT entries still get used.

\section{Tests and results}
\subsection{General setup}
The demonstration design runs on a SPEC board equipped with a FMC DIO 5-channel daughterboard.

Test signals go through the FMC daughterboard. The first LEMO connector on the daughterboard is configured as output and transmits an oscillating pattern. The next two LEMO connectors are inputs connected to TDC channels.

For measuring the FPGA temperature, a 1-wire digital thermometer is attached on top of the FPGA using kapton tape. Thermal paste improves conduction between the FPGA and the sensor.

The TDC core is configured with 2 channels (to enable differential measurements, see \S \ref{sec:dm}) and each delay line has 124 \verb!CARRY4! elements (496 taps).

To minimize variations of the timing properties between runs of the automated place and route tool and to maximize thermal coupling between each delay line and its online calibration oscillator, the design is floorplanned.

\begin{figure}[h!]
\centering
\includegraphics[width=1.6cm]{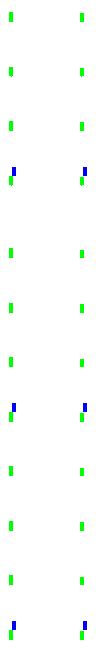}
\caption{Floorplan of the delay lines and ring oscillators in FPGA Editor.}
\label{fig:floorplan}
\end{figure}

The two delay lines from each channel are placed close to their respective IOBs. The ring oscillator components are placed in the \verb!SLICEX! columns just at the right of the delay lines, and spread evenly along the height of the delay lines. There is one ring oscillator per channel, which is made of many LUTs in series. This is illustrated by Figure \ref{fig:floorplan}, where the delay lines are colored green and the ring oscillators are blue (each blue block is a \verb!SLICEX! component containing a LUT belonging to one of the two ring oscillators).

\begin{figure}[h!]
\includegraphics[width=\columnwidth]{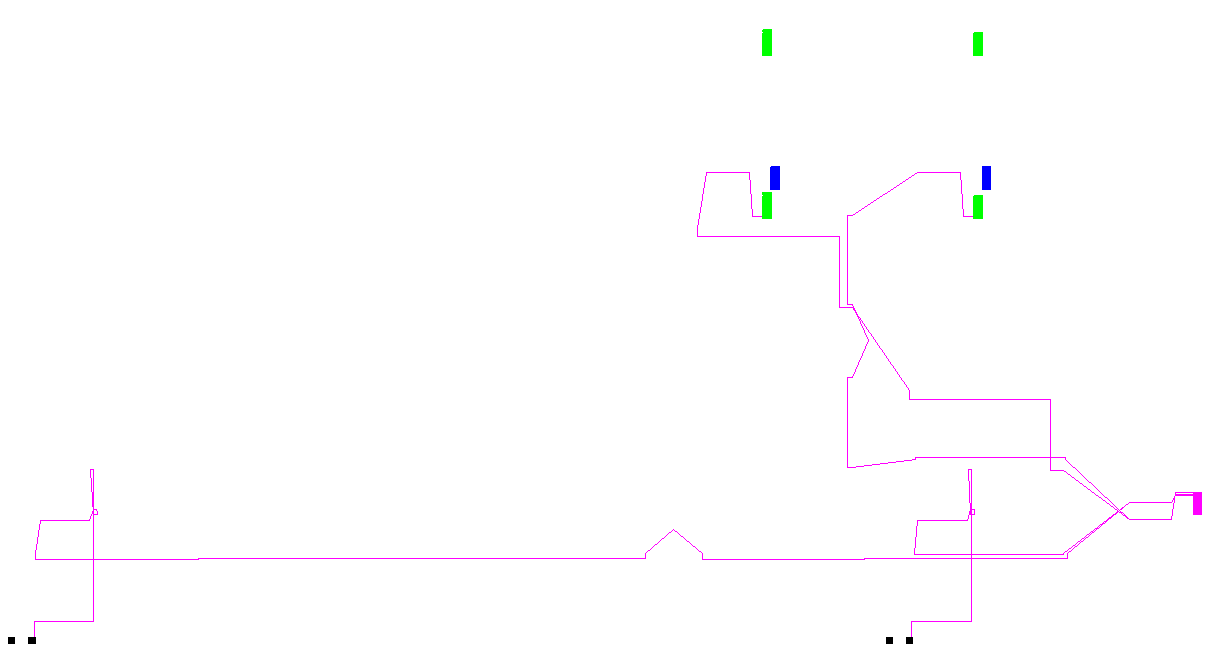}
\caption{Input signal path in FPGA Editor.}
\label{fig:inputpath}
\end{figure}

In the input signal path, there are one multiplexer and one inverter per channel. Everything is packed into one FPGA slice, which is also manually placed to minimize timing variations. The physical input signal path can be seen in Figure \ref{fig:inputpath}. The LVDS IOBs are represented in black, and the routing and the slice in pink.

A limitation of this TDC design is that it does not compensate for PVT variations in the input signal path elements.

\pagebreak

\section{Methods and results}

\subsection{Effect of temperature on ring oscillators}
\label{sec:rofreq}

\begin{figure}[h!]
\includegraphics[width=\columnwidth]{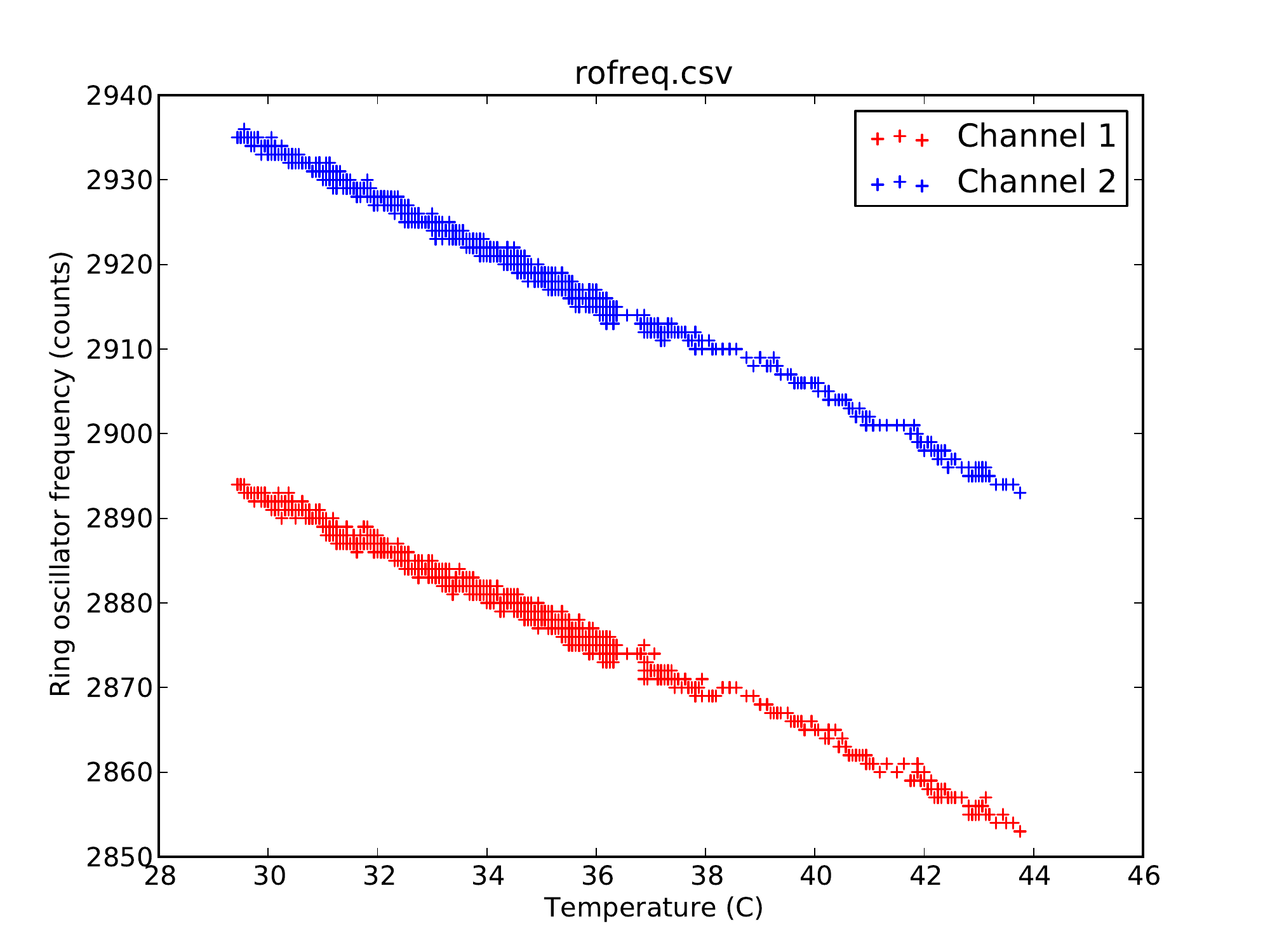}
\caption{Dependence of ring oscillator frequencies on temperature.}
\label{fig:rofreq}
\end{figure}

The purpose of this experiment is to examine how temperature affects propagation delays. We slowly heated the FPGA (so it remains in thermal equilibrium with the sensor) to obtain the plot of Figure \ref{fig:rofreq}.

The frequency values are directly reported from the TDC core, and are measured in cycles per frequency counter period.

As expected, the frequencies decrease linearly with the temperature, and the two channels follow a near-identical pattern. The variation is small: about 1.3\% for the 15\degree C difference. However, near the end of the delay line, a 1.3\% variation represents about 100ps, so it is important to compensate for the effects of temperature.

We suspect that the constant difference between the two channels is due to process variations across the different locations of the FPGA chip where the two ring oscillators are placed, and/or differences in routes chosen by the \verb!par! tool to implement the two oscillators.

\subsection{Startup calibration stability}
The startup calibration process relies on an asynchronous clock source which generates TDC events with a uniform random distribution within the system clock cycles. We wanted to verify that the process is deterministic enough.

\begin{figure}[h!]
\includegraphics[width=\columnwidth]{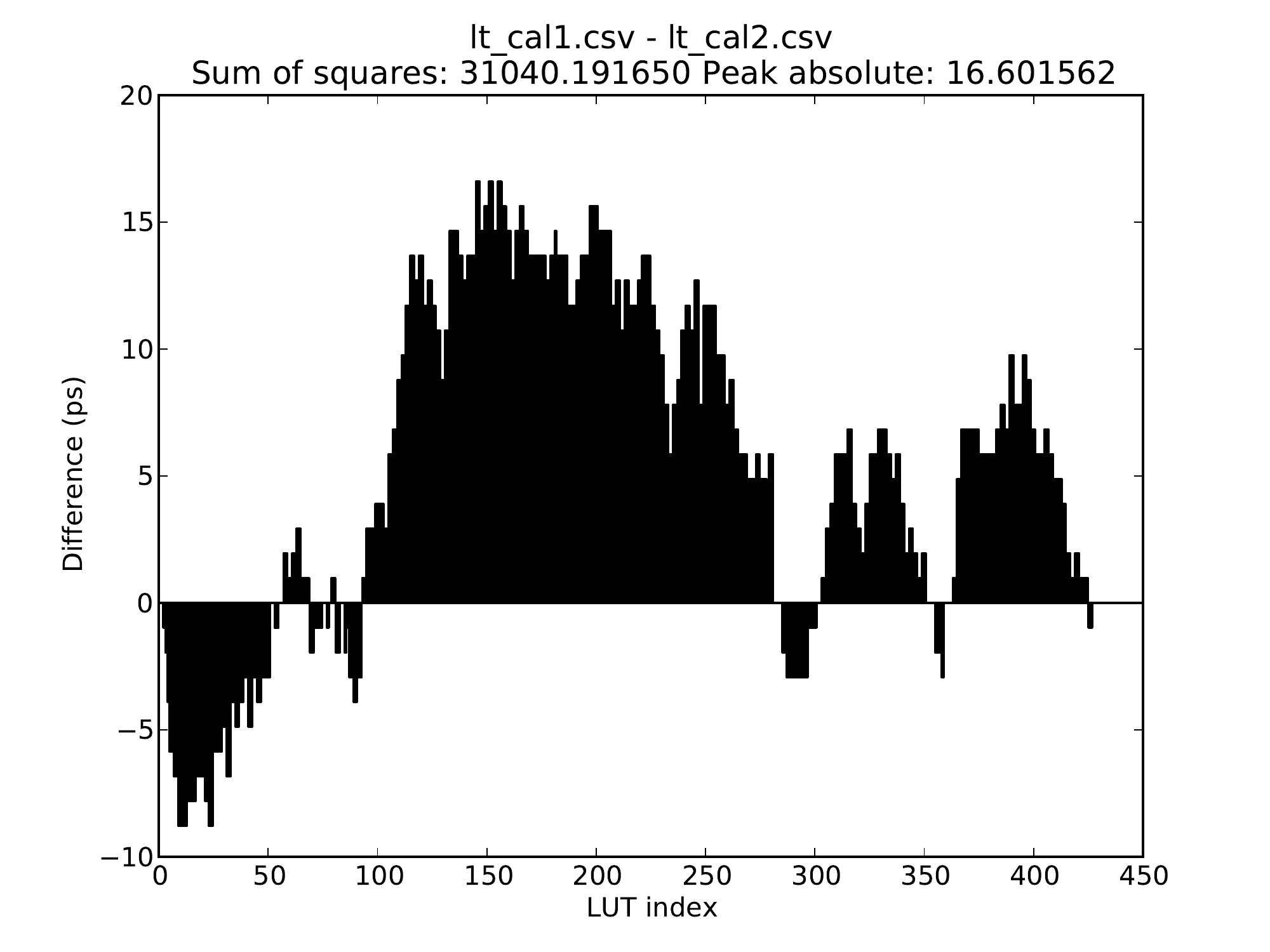}
\caption{Difference between the LUT contents from two startup calibrations at the same temperature.}
\label{fig:scs}
\end{figure}

With the FPGA in thermal equilibrium, we ran the startup calibration twice and compared the resulting LUT contents. The difference is plotted in Figure \ref{fig:scs}, and is small enough.

\subsection{Differential measurements}
\label{sec:dm}
The purpose of this test is to determine the precision of the system.

\begin{figure}[h!]
\includegraphics[width=\columnwidth]{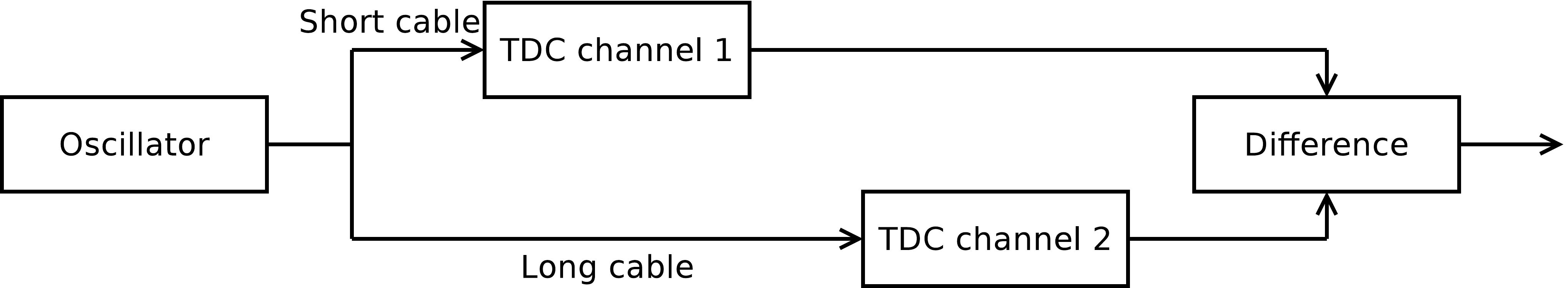}
\caption{Principle of differential measurements.}
\label{fig:dtdc}
\end{figure}

We connected the oscillator output of the FMC DIO card to a splitter feeding two cables of different lengths going to the two TDC channels. Those cables had propagation delays of approximately 2ns and 4ns. We then observed the difference between the two TDC timestamps, which is expected to remain constant (Figure \ref{fig:dtdc}). Since the oscillator is asynchronous to the system clock, the complete delay line can be covered and tested.

The advantage of this technique is that it is easy to set up and does not require expensive equipment. A limitation is that the result is not affected by common-mode noise of the input path to the delay line (Figure \ref{fig:inputpath}).

\begin{figure}[h!]
\includegraphics[width=\columnwidth]{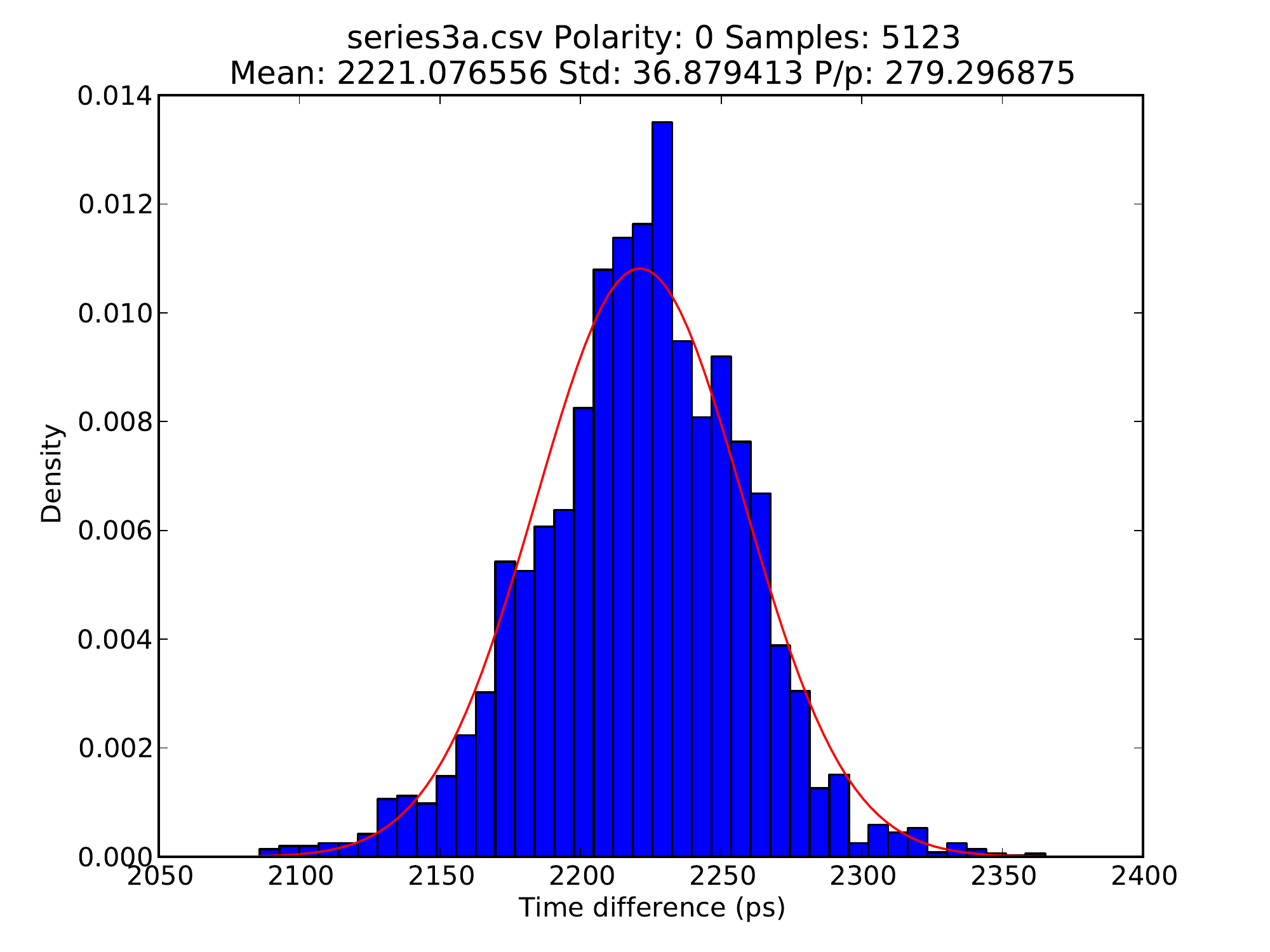}
\caption{Differential measurement results.}
\label{fig:mhist}
\end{figure}

We made the measurements at thermal equilibrium, with the sensor measuring 36.9375\degree C. The histogram of the results is shown in Figure \ref{fig:mhist}.

The results can be modeled with a Gaussian distribution having a mean of 2221ps (which is close to the 4ns-2ns difference in propagation times from the cables) and a standard deviation of 37ps. If we suppose that the jitter in each channel is independent and also has a Gaussian distribution, we can estimate that its standard deviation is 26ps. This means that for one channel, 95\% of the results are precise to $\pm$52ps.

\subsection{Temperature compensation}
Even though the influence of temperature is small (\S \ref{sec:rofreq}), we can still see the positive action of the online calibration.

\begin{figure}[h!]
\includegraphics[width=\columnwidth]{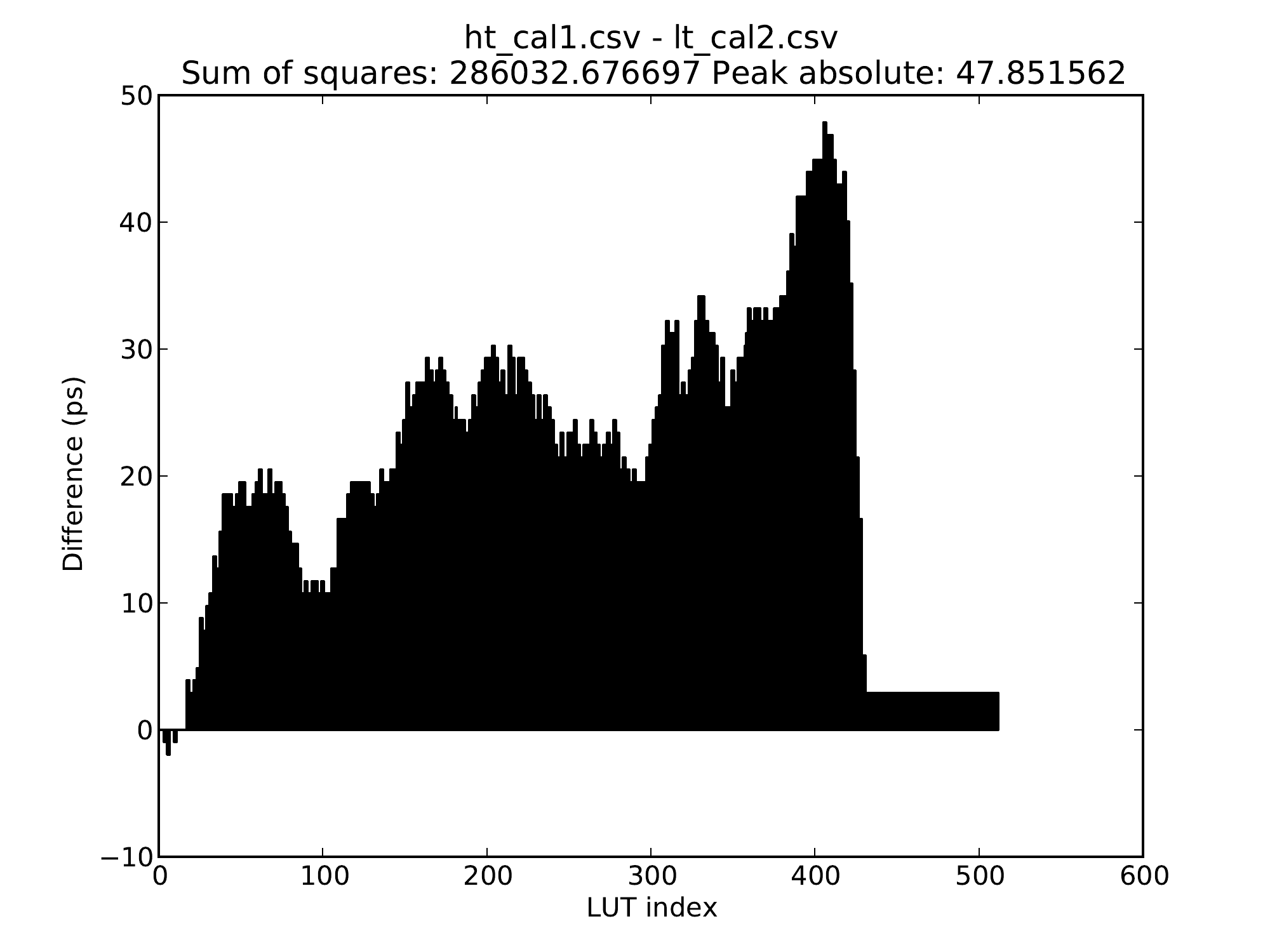}
\caption{Difference between the LUT contents from two startup calibrations at high and low temperatures.}
\label{fig:chtmlt}
\end{figure}

After calibrating at 37\degree C, we brought the temperature to 47.875\degree C, and ran startup calibration again. We observed a significant difference between the LUT contents (figure \ref{fig:chtmlt}).

\begin{figure}[h!]
\includegraphics[width=\columnwidth]{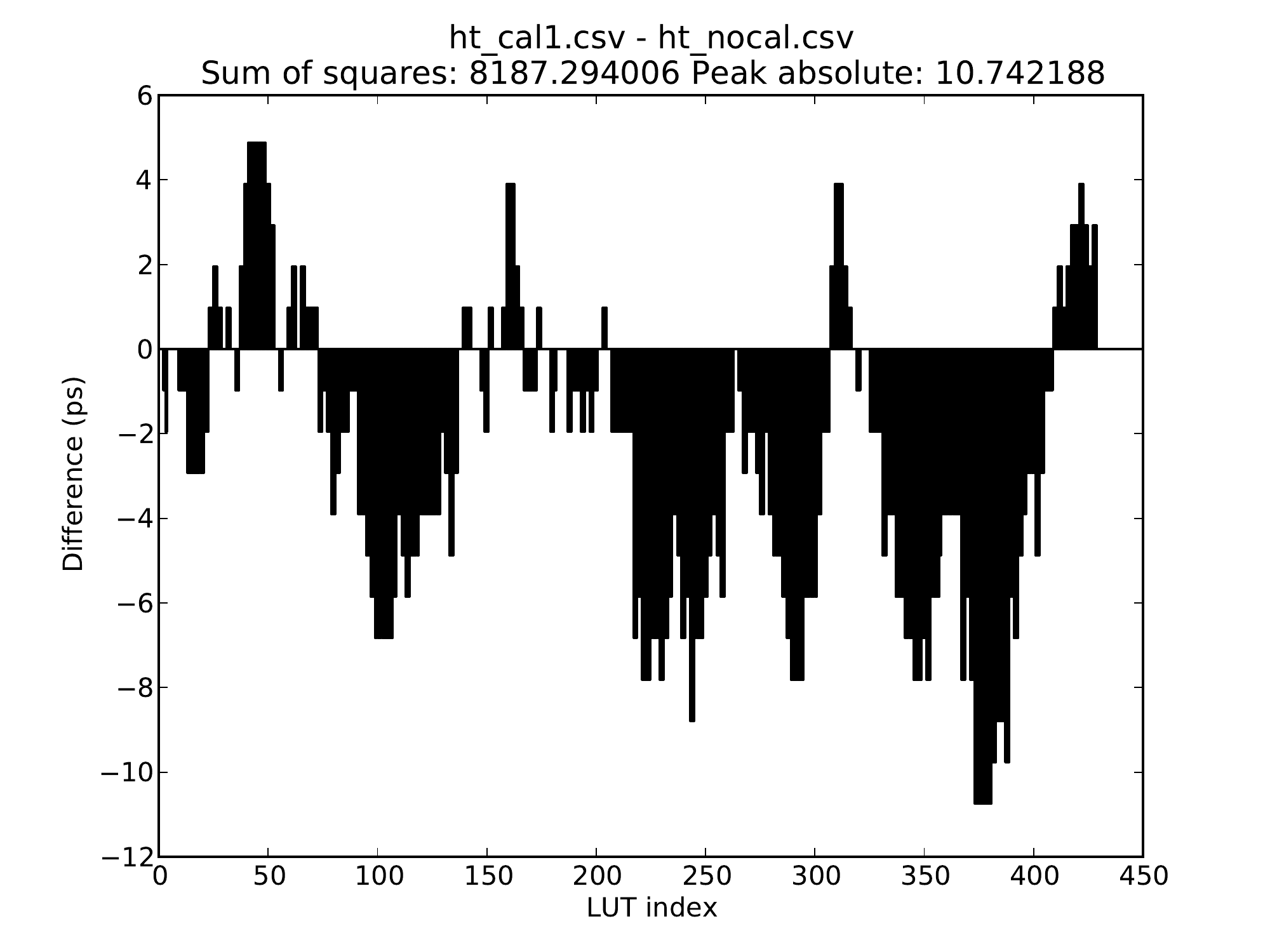}
\caption{Difference between the LUT contents from startup calibration and the values computed by online calibration.}
\label{fig:chtmht}
\end{figure}

The new LUT data are very close to what had been extrapolated from the 37\degree C data by the online calibration system (Figure \ref{fig:chtmht}). In fact, in this sample the difference is slightly smaller than what we had observed between two startup calibrations at the same temperature (Figure \ref{fig:scs}). This shows the good working of the online calibration system.

\section{Conclusions and future works}
The results of this experiment are very encouraging, as they show a precision better than that of many commercial TDC chips, but using a lower-cost FPGA. We can also potentially support dozens of channels in the XC6SLX45T of the SPEC board as the core uses little FPGA resources and also shares the calibration logic among all channels. Further, the latency of the core is low (6 cycles of the system clock) and the throughput high (for each channel, the dead time after an event is 3 cycles of the system clock, which can be brought down to 1 cycle without major architecture changes). This is also better than many commercial solutions.

There are however several areas of improvement.

First, more testing would be welcome, with many boards and FPGAs, with deliberate variations of the supply voltage, and within a wider temperature range.

Examining the startup calibration histograms reveals that almost half of the bin widths are zero. This is due to the particular propagation characteristics of carry chains, which are not the best solution for a delay line (their advantage, however, is that it is relatively easy to keep the exact same delays between runs of the place-and-route tool). It can make sense to use regular LUTs and/or general routing to implement the delay line instead, at the cost of increased design difficulty and reduced portability.

The startup calibration process could be improved (and made almost deterministic) by using as calibration signal a clock whose frequency is slightly different from the system clock. This way, the variations shown in Figure \ref{fig:scs} (which peak at almost 17ps) can be reduced or eliminated.

The carry chain is very long and this restricts its possible placements and compatibility with smaller FPGAs. Using LUTs and/or routing would also alleviate this problem.

If better precision is needed, multiple delay lines can work in parallel and their outputs combined, in order to average errors out.

Finally, the influence of the input path (Figure \ref{fig:inputpath}) was not thoroughly studied, even though we expect it to be minor.

\section{Acknowledgments}
This work was prepared under an agreement with and funded by CERN. The content of this paper does not necessarily reflect the position or the policy of CERN and no official endorsement should be inferred.

The author wishes to thank Javier Serrano and Tomasz Wlostowski for their valuable ideas and inputs.

\end{document}